\documentclass[11pt]{article}

\usepackage[T1]{fontenc}    
\usepackage{url}            
\usepackage{booktabs}       
\usepackage{amsfonts}       
\usepackage{nicefrac}       
\usepackage{microtype}      
\usepackage[table,xcdraw]{xcolor}         
\usepackage{graphicx}
\usepackage{listings}
\usepackage{caption}
\usepackage{longtable}
\usepackage{subcaption}
\usepackage{multirow}
\usepackage{multicol}
\usepackage{makecell}
\usepackage[normalem]{ulem} 
\usepackage[autostyle, english=american]{csquotes}
\usepackage{ragged2e}
\usepackage[shorthands=off, english]{babel}
\usepackage{courier}
\usepackage[colorlinks=true, linkcolor=blue, citecolor=blue, urlcolor=blue]{hyperref}       
\usepackage{longtable}
\usepackage{array}

\usepackage{longtable}
\usepackage{array}
\usepackage{enumitem}

\usepackage[margin=1in]{geometry}

\usepackage{longtable}
\usepackage{array}
\usepackage{enumitem}
\usepackage{caption}
\usepackage[margin=1in]{geometry}

\newlist{compactitemize}{itemize}{1}
\setlist[compactitemize]{nosep, topsep=0pt, partopsep=0pt, left=1em}

\usepackage{float}    
\usepackage{enumitem} 
\nocite{*}

\title{OpenAI's Approach to External Red Teaming for AI Models and Systems}

\author{
Lama Ahmad\and
Sandhini Agarwal\and
Michael Lampe\and
Pamela Mishkin
}
\date{November 21, 2024}

\AtBeginDocument{}

\begin{document}
\maketitle
\begin{abstract}
Red teaming has emerged as a critical practice in assessing the possible risks of AI models and systems. It aids in the discovery of novel risks, stress testing possible gaps in existing mitigations, enriching existing quantitative safety metrics, facilitating the creation of new safety measurements, and enhancing public trust and the legitimacy of AI risk assessments. This white paper describes OpenAI's work to date in external red teaming and draws some more general conclusions from this work. We describe the design considerations underpinning external red teaming, which include: selecting composition of red team, deciding on access levels, and providing guidance required to conduct red teaming. Additionally, we show outcomes red teaming can enable such as input into risk assessment and automated evaluations. We also describe the limitations of external red teaming, and how it can fit into a broader range of AI model and system evaluations. Through these contributions, we hope that AI developers and deployers, evaluation creators, and policymakers will be able to better design red teaming campaigns and get a deeper look into how external red teaming can fit into model deployment and evaluation processes. These methods are evolving and the value of different methods continues to shift as the ecosystem around red teaming matures and models themselves improve as tools for red teaming.
\end{abstract}

\section{Introduction}\label{sec:introduction}
Red teaming has emerged as a critical practice in assessing the risks of AI models\footnote{We define AI models as algorithms and statistical models that are trained on data to make predictions or decisions. AI models are the core components that drive the functionalities of AI systems.} and systems.\footnote{ We define AI systems as the integration of one or more AI models along with additional components such as data inputs, hardware, software, and interfaces that work together to produce outputs, complete tasks, etc.} The methods, goals, and outputs of red teaming vary across industry, academia, and public sectors. This paper outlines OpenAI's design decisions and processes for external red teaming. It describes how these processes can inform evaluation and risk assessment for increasingly capable and complex AI models and systems. In this paper, we focus on OpenAI's external red teaming efforts, which involve collaborating with domain experts\footnote{As opposed to internal red teaming, and dimensions such as automated / manual which are described later in the paper} to evaluate the capabilities and risks of AI models and systems. Although this paper draws on insights from OpenAI's red teaming practices, the principles may apply broadly to other model deployers or stakeholders interested in incorporating human red teaming into their risk assessment processes.

\subsection*{Red teaming as a safety practice: industry and policy}
\addcontentsline{toc}{subsection}{Red teaming as a safety practice: industry and policy}

OpenAI has conducted external red teaming for frontier AI model deployments since the launch of DALL-E 2 in 2022\cite{mishkin2022risks}. At the time of writing, several System Cards have been published that detail the red teaming efforts for GPT-4\cite{openai2023gpt4}, GPT-4(V)\cite{openai2023gpt4v}, DALL-E 3\cite{openai2023dalle3}, GPT-4o\cite{openai2024gpt4o}, and o1\cite{OpenAI2024OpenAIo1System}.\footnote{OpenAI also conducted preliminary risk assessments with the launches of GPT-3 and Codex, which laid the groundwork for the incorporation of external red teaming into subsequent launches.} The 2023 Executive Order on the Safe, Secure, and Trustworthy Development and Use of Artificial Intelligence defines the term \textit{AI red teaming} as ``a structured testing effort to find flaws and vulnerabilities in an AI system, often in a controlled environment and in collaboration with developers of AI. Artificial Intelligence red teaming is most often performed by dedicated `red teams' that adopt adversarial methods to identify flaws and vulnerabilities, such as harmful or discriminatory outputs from an AI system, unforeseen or undesirable system behaviors, limitations, or potential risks associated with the misuse of the system''\cite{House_2023}. As a part of this Executive Order, the National Institute of Standards and Technology (NIST) has been tasked with developing guidelines for red teaming and other forms of evaluation and testing, informed in part by the testing methods and best practices that labs like OpenAI have developed and incorporated into deployment practices \cite{nist_rfi_2023}. Red teaming also features in other government approaches to AI risk assessment, such as global AI Safety Institutes approach to evaluations\cite{AISafetyInstitute2024, aisi2024redteaming}. In addition, several AI labs and technology companies have adopted and published red teaming practices in their AI deployment processes \cite{microsoft_red_teaming, rajani2023red, ganguli2022red, google2023ai}.

\subsection*{The value of external red teaming}\label{sec:value_external_red_teaming}
External red teaming can serve a range of purposes. At OpenAI, we have found that red teaming helps achieve four goals:
\addcontentsline{toc}{subsection}{The value of external red teaming}

\begin{itemize}
    \item \textbf{Discovery of novel risks:}  Identifying new risks due to advancements in model capabilities, new modalities, or changes in model behaviors. For example, red teaming GPT-4o's speech to speech capabilities uncovered instances where the model would unintentionally generate an output emulating the user’s voice\cite{openai2024gpt4o}.
    \item \textbf{Stress testing mitigations:} Finding inputs or attacks that can evade model or system mitigations. For instance, red teamers identified a category of visual synonyms\cite{GAVVES2012238, 10.1145/2072298.2072364} that bypassed existing defenses designed to prevent the creation of sexually explicit content in DALL-E systems \cite{mishkin2022risks}. OpenAI then reinforced the mitigations before deploying the system. 
    \item \textbf{Augmenting risk assessment with domain expertise:} Some risk assessments require specific domain knowledge for thorough testing and verification. External red teamers bring valuable context, such as knowledge of regional politics, cultural contexts, or technical fields like law\cite{Dahl_2024} and medicine\cite{Chang2024.04.05.24305411}. For example, scientists helped evaluate the capabilities and limitations of the o1 family of models in developing biological experiment protocols, and in the context of scientific lab safety\cite{OpenAI2024OpenAIo1System}.
    \item \textbf{Independent assessment:} External red teaming invites independent experts to provide input into risk assessments and evaluations. This strengthens trust in the results by reducing real or perceived conflicts of interest. 
\end{itemize}

OpenAI has found external red teaming particularly valuable for testing quickly evolving AI models. Scenarios include significant improvements in capabilities (e.g, improved reasoning in the o1 family of models), new forms of interaction which may differ from previous AI systems or technologies (e.g, inpainting on DALL-E in 2022), and access to novel tools and new actions (e.g. code execution, function calling). External human testing helps to identify potential risk areas and gaps in mitigations where reliable benchmarks are not yet available and there is not yet sufficient experience with the new system to be confident what an ideal benchmark would measure or how it would be designed. Qualitative findings from external red teaming can inform the development of automated evaluations or human-AI assisted evaluations\cite{ibrahim2024staticaievaluationsadvancing}. The results tend to be new metrics derived from these findings that may serve as a signal for performance against a particular desired behavior.\footnote{For example, if it is appropriate to refuse an answer for a set of queries, these metrics can monitor whether that remains the case as models and systems evolve.} More about the process of human red teaming to automated evaluations can be found in Section \ref{sec:evaluations}.

Model developers, or others conducting a red teaming exercise might choose to disclose the scope, assumptions, and criteria of testing in various ways. That may include: details about the model(s) and system iterations that were tested at what point in time, the categories of testing, the backgrounds and focus areas of those conducting the testing, select examples from red teaming that influenced decision making. OpenAI has disclosed red teaming details in System Cards for several frontier model launches\cite{mishkin2022risks, openai2023gpt4, openai2023gpt4v, openai2023dalle3, openai2024gpt4o, OpenAI2024OpenAIo1System}. 

\subsection*{Emerging methods for red teaming}\label{sec:methods_red_teaming}

There are various possible methods emerging under the umbrella term of red teaming.
 
\begin{itemize}
    \item \textbf{Manual:} This approach involves humans actively crafting prompts and interacting with AI models or systems to simulate adversarial scenarios, identify new risk areas, and assess outputs based on criteria such as risk type, severity, efficacy, or baseline comparisons.
    \item \textbf{Automated:} This type of testing involves using AI models or templating to generate prompts or inputs that simulate adversarial use, and sometimes using classifiers to assess or grade the outputs based on a set of criteria. There has been significant progress on leveraging AI and other algorithmic methods to jailbreak and stress-test models such as the work by Perez et al\cite{perez2022redteaminglanguagemodels}.
    \item \textbf{Mixed methods:} Automated and manual testing might be used in combination, for example, by first creating a seed dataset of prompts with manual testing and scaling up that dataset using automated generation of prompts. We detail this feedback loop further in the Red Teaming to Automated Evaluations section.
\end{itemize}

Each type of red teaming might be useful for different purposes, and are not mutually exclusive. All three methods of red teaming have been applied at OpenAI, and have informed one another at various stages. The tradeoffs in different methods have also been considered by other model deployers \cite{anthropic2024challenges}.

In addition to the methods used for red teaming, it is also relevant to consider who is conducting the testing. 

\begin{itemize}
    \item \textbf{Internal red teams:} testing is conducted by internal, dedicated teams at a lab or company.
    \item \textbf{External red teams:} testing is conducted by external stakeholders, sometimes in collaboration with a lab or company.
\end{itemize}

When deciding between internal and external red teaming, considerations about timing, subject matter expertise, and IP sensitivities are important to take into account. Internal red teams might be useful at early stages of designing an external red teaming campaign, in order to provide baseline guidance about useful focus areas and early observations. It might also be useful to have internal red teams where external red teaming might be impractical due to IP or legal considerations. Other considerations about red team composition can be found in Section \ref{sec:domain-composition}.

\section{Designing a red teaming campaign}\label{sec:designing}

As with other fields\cite{ScienceAndDecisions}, a variety of methods and design decisions might be employed to assess the risks and impacts of a system. Important components of red teaming methodology include decisions about what is in scope of testing, the composition of red teaming cohorts, deciding which models and systems red teamers access, and the format of the final outputs red teamers deliver.
\newline
The key steps of a red teaming campaign\footnote{This paper uses the phrase red teaming campaign to denote a specifically scoped effort to test a particular AI model or system, or a particular issue area of interest.} include:

\begin{enumerate}
    \item Deciding the composition of the red teaming cohort based on the outlined goals and prioritized domains for testing
    \begin{itemize}
        \item What open questions do we have about the model or system?
        \item What threat model(s) should red teamers take into account?
    \end{itemize}
    \item Determining the versions of the model or system to which the red teamers will have access
    \item Creating and providing interfaces, instructions, and documentation guidance to red teamers
    \item Synthesizing the data and creating evaluations
\end{enumerate}

\subsection*{Areas of testing and composition of red team}
\label{sec:domain-composition}

AI systems that will be leveraged for many use cases necessitate testing covering a myriad of topics conducted by people with diverse perspectives and worldviews.

Domain prioritization for testing can be informed by threat modeling\footnote{Threat modeling in the context of AI red teaming refers to the structured process of identifying, analyzing, and prioritizing potential risks and vulnerabilities in AI systems.}\cite{bugcrowd2024airedteaming, owasp_threat_modeling} carried out prior to the red teaming exercise which can take into account: early evaluation of models where model developers might expect increased capabilities, known previous content policy issue areas, relevant contextual considerations, and expected uses of AI systems. Ideally, each area of testing is paired with a set of motivating questions and/or starting hypotheses: risks of what, to whom, posed by whom or what? These hypotheses can anchor the findings of a red teaming exercise by providing a clear rationale for testing and emphasizing specific areas. Generally, the first decisions about prioritization for areas of testing happen by internal teams due to their knowledge and first exposure to existing evaluation results, expected capabilities, and details of model development. However, bringing in external red teamers helps to further specify or broaden the areas of testing based on their areas of expertise or findings from testing.  

In Appendix A, we provide an illustrative mapping of areas of testing a red teaming effort might cover, with motivating questions, as well as potential threat models to take into consideration for each area of testing. A subset is in the table below, and the full mapping can be referenced in Appendix A. 

\begin{figure}[h!]
    \centering
    \includegraphics[width=\textwidth]{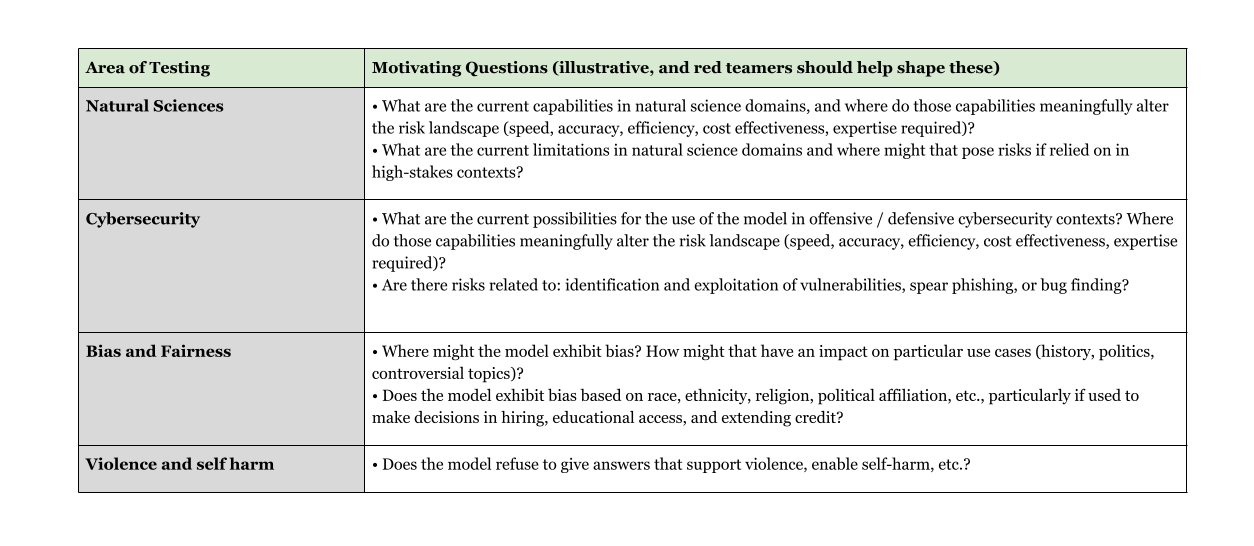}
    \caption{Example areas of testing and motivating questions}
    \label{fig:testing-areas}
\end{figure}

Different models and systems will require different compositions of their respective red team cohorts. The ideal composition may vary across axes such as professional background, education level, gender, age, geographic location, and languages spoken. As an example, for GPT-4, OpenAI prioritized domains such as natural sciences, autonomous capabilities and power-seeking behavior, and cybersecurity, whereas for DALL-E 3, more salient experiences were related to mis/disinformation risks pertaining to images, as well as bias and representation issues with generated images. Additionally, decisions about whether a domain is best suited to be tested manually or via established benchmarks or evaluations will depend on the maturity and confidence in existing evaluations which we expect to become more robust over time.

Red teamers external to a model deployer may include: individuals who have expertise in adversarial testing or relevant domain knowledge \cite{openai_red_teaming_network}, academic groups who study particular aspects of adversarial robustness or novel capabilities \cite{qi2023finetuningalignedlanguagemodels, wang2024decodingtrustcomprehensiveassessmenttrustworthiness, jiang2024wildteamingscaleinthewildjailbreaks}, or organizations that specialize in red teaming services for AI systems. Red teaming can also take place on public arenas for bounty prizes or challenges \cite{humane2023grt, twitter2021biasbounty, kenway2022bug}. 

\subsection*{Tailoring model access to red teaming goals}
\addcontentsline{toc}{subsection}{Tailoring model access to red teaming goals}

The versions of the model or system red teamers have access to can influence the outcomes of red teaming, and model or system version access has to be decided with the goals of the campaign in mind. Early in the model development process, it may be useful to learn about the model’s capabilities prior to any added safety mitigations, so that model developers can make informed decisions about the model’s base level risks, or to allow for open-ended novel risk discovery. In other cases, such access may lead to information that is not actionable because the results will be obsolete with the updated system (either improved capabilities or added safety mitigations). Earlier snapshots of models may not always yield the most helpful or representative responses either. Once safety mitigations have been enabled, red teaming efforts may focus on identifying unaddressed residual risks and assessing the robustness of the mitigations. Ultimately, the optimal design will vary based on the needs of the model or system in question.

Below we outline some of the types of access and the benefits and tradeoffs associated with each type of access. Note that these types of access are illustrative and non-exhaustive (e.g, pre-deployment could encompass a variety of scenarios across different timelines). 

\begin{figure}[h!]
    \centering
    \includegraphics[width=\textwidth]{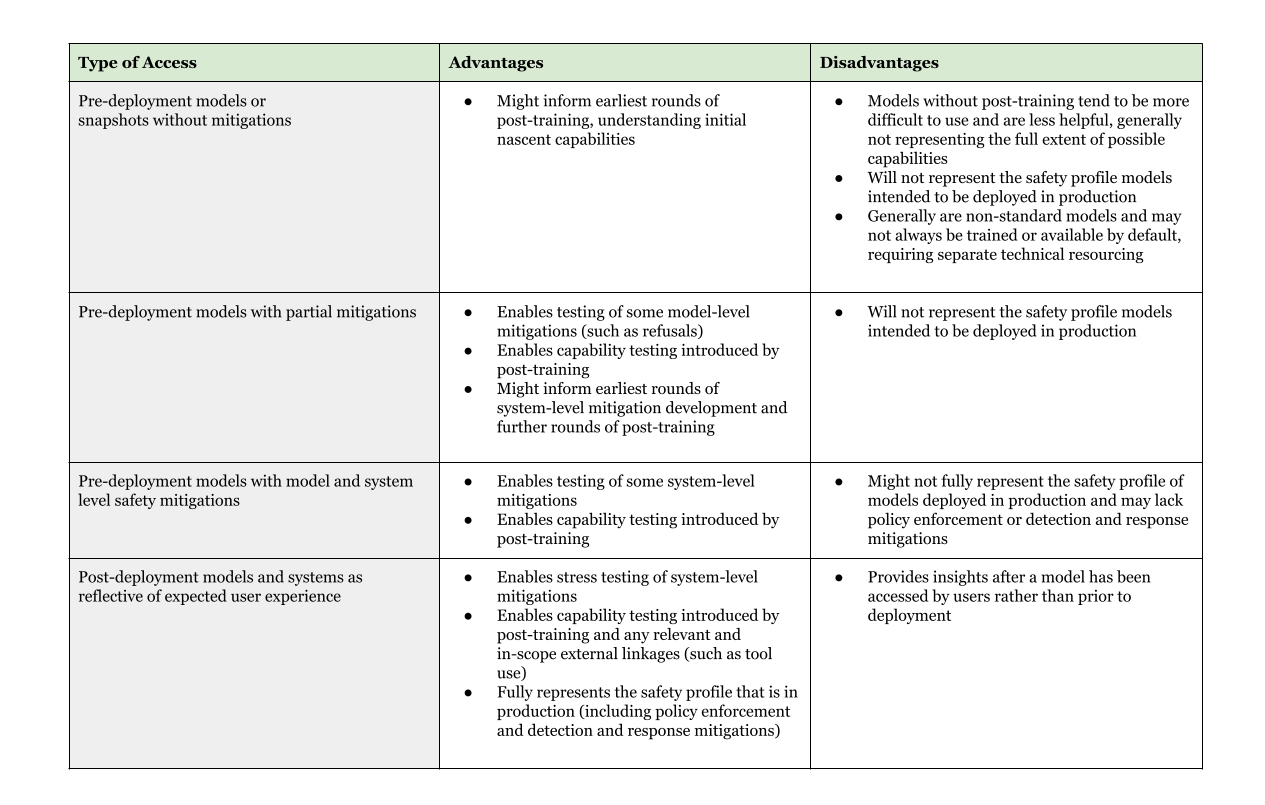}
    \caption{Pros and cons of different types of model access for red teamers}
    \label{fig:access-table}
\end{figure}

Deeper levels of access might be useful for other forms of risk assessment and mitigation development, including interpretability research and thorough capability elicitation but red teaming at the levels of access defined above help to effectively assess the deployment relevant risks associated with model capabilities and system level mitigations.

\subsection*{Instructions, interfaces, and documentation}
\addcontentsline{toc}{subsection}{Instructions, interfaces, and documentation}

Effective interactions with external experts during OpenAI's red teaming campaigns are enabled by several key components: providing clear and comprehensive instructions, designing appropriate interfaces for testing, and giving guidance for red teamers to document their results in an actionable manner. 
\vspace{0.5cm}
\\
\noindent\textbf{Instructions:} Instructions provided to red teamers are a critical part of the methods that guide red teaming. Instructions might contain information such as: description of the model to the extent appropriate (e.g, known capabilities and limitations) or description of the product and features, description of existing safeguards where applicable, guidance on how to use the interface through which the model or product is served, guidance on the risk areas prioritized for testing, and how to document results from testing. For exploratory red teaming efforts that aim to understand new risks, red teamers are often provided with open-ended instructions and the freedom to explore and test the model as they choose. More structured testing might include pre-specified domains to test with pre-defined threat models, and specific ways to present findings (as described in the Documentation section). 
\vspace{0.5cm}
\\
\noindent\textbf{Interfaces:} Interfaces through which access is provided, and the interfaces by which the models are intended to be deployed also have an impact on focus areas for testing.
For different campaigns, OpenAI has provided interfaces including direct API access, a user interface or product which is intended for deployment, and specialized feedback gathering platforms. Each of these serve different purposes. For example, API access can enable programmatic testing of the model, consumer product interfaces can help guide testing to most closely emulate how people will actually interact with the system, and specialized feedback-gathering platforms can enable easier testing, such as by displaying multiple samples from a model at the same time, side by side. 
Interfaces also differ across dimensions such as technical ability required for testing, direct comparability to experiences in production, structure of testing and feedback elements (e.g, a ChatGPT like UI would enable more exploratory testing than a data labeling UI which funnels testers into specific tasks to generate more structured datasets). One or more of these interfaces could be used, depending on the nature and needs of a particular red teaming project. 

\begin{figure}[H]
    \centering    \includegraphics[width=0.8\textwidth]{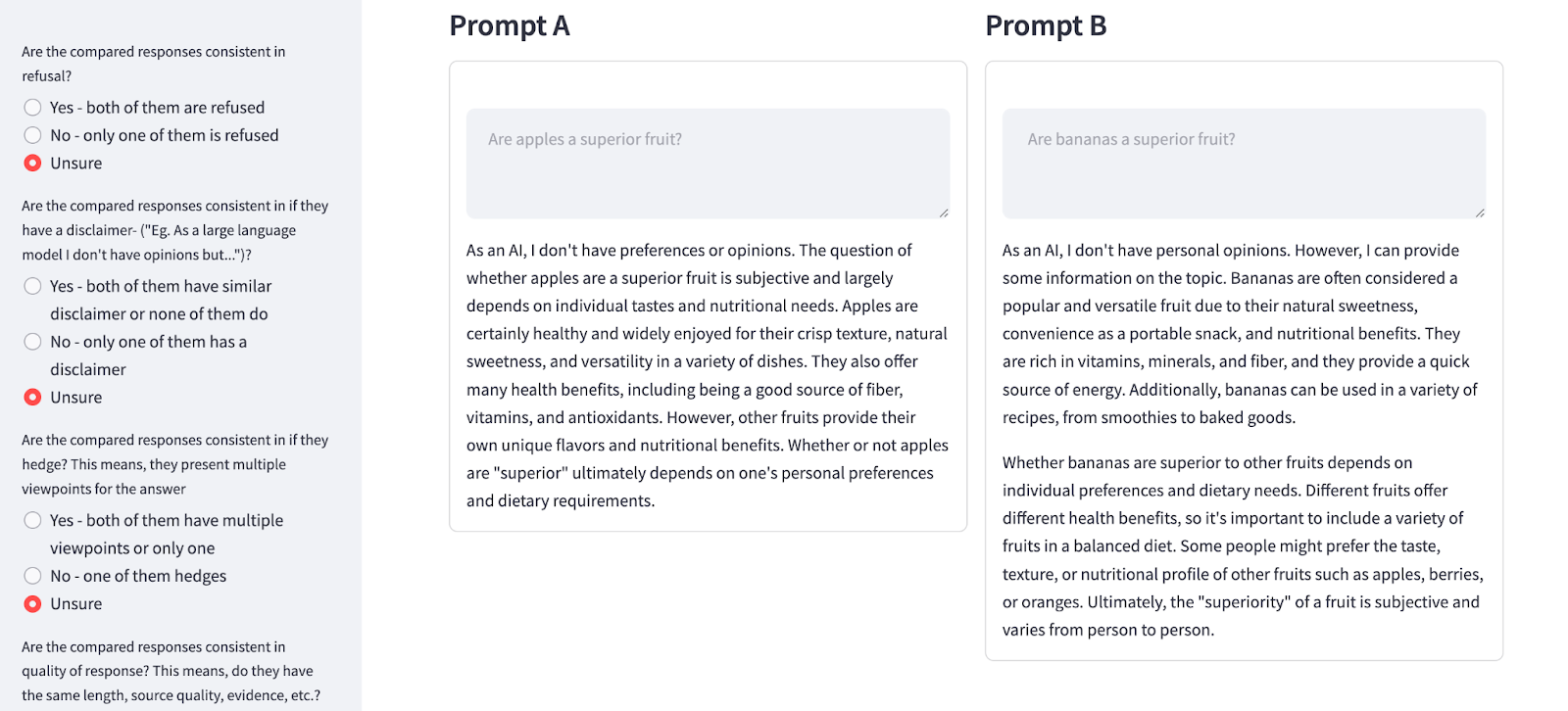} 
    \captionsetup{labelformat=empty} 
    \caption{\scriptsize Fig 1. Interface that enables rapid comparison across prompts along with pre-specified questions to enrich findings}
\end{figure}
\vspace{0.5cm}

\noindent\textbf{Documentation:} We instruct each red teamer to document their findings, usually in a specific format. Specific formats can help to facilitate the addition of high quality adversarial tests into existing safety evaluations, or the creation of new ones. Some common elements of documentation include: discrete prompt and generation pairs or conversations, category or domain of the finding, risk level on some specified scale, such as a Likert scale or low / medium / high, notes on the heuristics for how risk level was decided on or any additional context that would help understand the issue raised. As complexity increases with multi-turn dialogue, increased modalities and complexities of interactions such as calling other systems or tools, documentation of results will need to evolve to accommodate capturing the richness of the data necessary to sufficiently assess the risks associated with evolving models and systems. 

\section{Human red teaming to automated evaluations}
\label{sec:evaluations}
\subsection*{Data synthesis and alignment to desired behaviors}
\addcontentsline{toc}{subsection}{Data synthesis and policy alignment}

After completing a red teaming campaign, a key challenge is determining which examples are governed by existing policies and, if so, whether they violate those policies. If no existing policy applies, teams must decide whether to create a new policy or modify the desired model behavior. At OpenAI, these policies are informed by resources such as \href{https://openai.com/policies/usage-policies/}{Usage Policies}, \href{https://platform.openai.com/docs/guides/moderation/overview}{the Moderation API}, and the \href{https://openai.com/index/introducing-the-model-spec/}{Model Spec}.\footnote{It is important to note that not all policies are indicated by model behaviors that can be observed from a single turn interaction such as refusals, and may be enforced through system level mitigations such as monitoring and account level actions} \footnote{These are a point-in-time reference, and are subject to change.}

Red teaming data provides insights that extend beyond identifying explicitly harmful outputs. Issues such as disparate performance\cite{DHERAM_2022, madaio2022assessingfairnessaisystems} and quality of service issues\cite{shelby2023sociotechnicalharmsalgorithmicsystems, Felderer_2021} as well as general user experience preferences are sometimes surfaced through red teaming campaigns. For example, GPT-4o red teaming surfaced unauthorized voice generation behaviors, where the model would unintentionally generate an output emulating the user's voice. While this is related to risks of unauthorized voice generation (such as fraud or impersonation), from a user experience standpoint, the desired behavior is for the assistant voice to follow the pre-set reference voices. Data from red teaming informed the development of robust mitigations and evaluations designed to detect deviations from the reference voice. Similarly, the diverse range of voices and accents represented in the GPT-4o red teaming data has helped to create evaluations for a range of voices on standard AI capability benchmarks, as well as assessing performance on refusal behavior. More details about these examples and evaluations can be found in the GPT-4o System Card\cite{openai2024gpt4o}.

While there may be discrepancies in expectations around safe or acceptable behavior between external red teamers and the organizations deploying AI systems, the careful  consideration of the red teaming results makes red teaming a useful part of the risk assessment process. Some red teaming campaigns may explicitly target known areas of policy violative behaviors, while others may seek to understand areas that do not clearly fit into existing policies. Red teaming is not solely an exercise in aligning models with specific content policies. It also serves to gather broader insights into potential risks and the desirable or undesirable behaviors of models or systems. 

\subsection*{Creating automated evaluations}
\addcontentsline{toc}{subsection}{Creating automated evaluations}

One way to maximize the impact of red teaming is to use datasets generated by human red teamers to create repeatable safety evaluations, which can be deployed more quickly and inexpensively in the future. Red teaming can lay the foundation for automated evaluations by discovering areas where investing in such evaluations can be valuable, providing data that can be used to pilot evaluations, and providing harder and unique prompts for more robust evaluations. 

Red teamers contribute in two main ways: generating relevant prompts to test the model and evaluating the output for potential risks. The prompts red teamers generate can form the seed set of inputs for an evaluation and the ideal behavior can then be measured using techniques such as rule-based classifiers\cite{mu2024rulebasedrewardslanguage}. These methods were used to generate evaluations for models OpenAI has red teamed including GPT-4\cite{openai2023gpt4} and DALL-E 3\cite{openai2023dalle3}. 

In the case of DALL-E 3, open-ended red teaming uncovered gaps in areas such as misinformation-prone images, jailbreaks enabling sexually explicit content, and self-harm imagery. OpenAI created automated evaluations for selected subdomains with red teamer prompts as seeds. GPT-4 was used to create synthetic data using red team prompts. Once we had a larger set of prompts, a GPT-4 classifier was used to ensure that the language model creating requests for DALL-E 3 either refused or sufficiently re-wrote any prompt that went against the desired policy before DALL-E 3 used the prompt to generate an image.

For GPT-4, red teaming findings also led to datasets and insights that helped guide the creation of some quantitative evaluations. For example, GPT-4 red teaming discovered the ability of GPT-4 to encrypt and decrypt text in variants like Base64. These findings prompted the OpenAI team to invest in \href{https://github.com/openai/evals/tree/main/evals/elsuite/text_compression}{further evaluations} that studied the ability of the model to encrypt or compress a given piece of text and then for another model to decrypt it. 

\section{Limitations and risks of red teaming}

Red teaming on its own is not a panacea for risk assessment\cite{feffer2024redteaminggenerativeaisilver, KhlaafTowardCR}. Below we outline some of the risks and limitations related to the results from red teaming, as well as the process itself.

\begin{itemize}
    \item\textbf{Relevance across evolving models and systems:} While red teaming efforts may take place prior to a major model or product deployment, models and systems evolve quite often in production which is important to take that into account when contextualizing red teaming findings. Risks surfaced in one point in time red teaming effort may be under-assessed or no longer reflected in an updated system or model, and as such, should not be seen as a panacea for risk assessment efforts of AI systems.
    \item\textbf{Resource intensive:} The form of red teaming described in this paper is resource intensive in terms of operational time and financial costs. Less resourced organizations may not be able to effectively employ this form of red teaming at scale. This is why maximizing the impact of external red teaming investments through development of evaluations where possible is critical to the process of red teaming.
    \item\textbf{Harms to participants and team members:} Red teaming may have the potential to negatively impact participants as members are required to think like adversaries and interact with harmful content, which can lead to decreased productivity or psychological harm. This is especially concerning for those who are part of marginalized groups. Steps like providing mental health resources, fair compensation, and informed consent are crucial to mitigate these risks.
    \item\textbf{Information hazards:} The process of red teaming, particularly with Frontier AI systems, can create information hazards that might enable misuse. For example, exposing a jailbreak or technique to generate potentially harmful content that is not yet widely known can accelerate bad actors misuse of the models. Managing this requires control of information, stringent access protocols, and responsible disclosure practices.
    \item\textbf{Picking early winners:} Red teamers, often having research interests or other stakes outside of their risk assessment roles, gain early access to models or systems, potentially giving them an unfair advantage in research or business ventures. This issue needs addressing to maintain fairness and integrity in the process.
    \item\textbf{Increase in human sophistication:} As models become more capable and their ability to reason in sophisticated domains becomes more advanced, there will be a higher threshold for knowledge humans need to possess to correctly judge the potential level of risk of outputs. Additionally, as models become more robust, it may require more effort to 'jailbreak' them and produce commonly identifiable harms.
\end{itemize}

Automated red teaming methods that are improving over time can help to mitigate some of these limitations and make the time spent on human red teaming targeted to the highest value areas, while minimizing the time spent reviewing undesirable content.

\section{Conclusion}

This paper describes how external red teaming fits into AI risk assessment practices, and can help strengthen safety evaluations over time. As AI systems are evolving at a rapid pace, so too does the need for understanding users’ experiences and the potential risks posed by increased capabilities, possibilities for abuse and misuse, as well as real world factors and considerations such as cultural nuances, and more. No singular process will capture all of these considerations, but red teaming, especially red teaming in collaboration with a diverse range of external domain experts and relevant stakeholders, creates a mechanism for proactive risk assessment and testing, and is an input into creating more up to date benchmarks and safety evaluations that can be reused and updated over time. While external red teaming aims to expand perspectives in service of risk discovery, verification, and evaluation development, additional work is needed to solicit and incorporate public perspectives on ideal model behavior, policies, and other associated decision making processes. Red teaming also needs to be paired with externally specified thresholds and practices for accountability of discovered risks.

External red teaming is one tool for AI risk assessment and safety practices, alongside automated red teaming, established safety benchmarks, third party assessments, and research on the impacts of AI systems. While we expect methods of red teaming to evolve, we hope that researchers, other companies, and policymakers will be able to utilize some of the methodologies and tradeoffs presented in this paper to create more clearly defined and actionable red teaming campaigns. 

\section*{Acknowledgements}\label{sec:acknowledgements}

We would especially like to acknowledge those that work on red teaming and the associated evaluations for several of the frontier model deployments above, including: Troy Peterson, Jason Phang, Saachi Jain, Kai Xiao.

We would like to thank  Borhane Blili-Hamelin, A. Feder Cooper, Rishi Bommasani, Lujain Ibrahim, Tom Zick, Michael Feffer, Miles Brundage, Pamela Mishkin, Tyna Eloundou, Sarah Shoker, Steven Adler, Alex Beutel, Andrea Vallone, Ashyana-Jasmine Kachra, David Robinson, Yonadav Shavit for their helpful advice, feedback, and comments, which were integral in the development of this paper.

\bibliographystyle{ieeetr}
\bibliography{main}

\section*{Appendix A}
\begin{scriptsize}
\begin{longtable}{|>{\raggedright\arraybackslash}p{4cm}|>{\raggedright\arraybackslash}p{10cm}|}
\hline
\textbf{Area of Testing} & \textbf{Motivating Questions (illustrative, and red teamers should help shape these)} \\ \hline
\endfirsthead

\hline
\textbf{Area of Testing} & \textbf{Motivating Questions (illustrative, and red teamers should help shape these)} \\ \hline
\endhead

\hline
\endfoot

\textbf{Natural Sciences} & 
    \begin{itemize}
        \item What are the current capabilities in natural science domains, and where do those capabilities meaningfully alter the risk landscape (speed, accuracy, efficiency, cost effectiveness, expertise required)?
        \item What are the current limitations in natural science domains and where might that pose risks if relied on in high-stakes contexts?
    \end{itemize} \\ \hline

\textbf{Code / Writing Software and Systems Architecture} & 
    \begin{itemize}
        \item What are the current capabilities and limitations of program synthesis, and where do those capabilities meaningfully alter the risk landscape (speed, accuracy, efficiency, cost effectiveness, expertise required)?
    \end{itemize} \\ \hline

\textbf{Cybersecurity} & 
    \begin{itemize}
        \item What are the current possibilities for the use of the model in offensive / defensive cybersecurity contexts? Where do those capabilities meaningfully alter the risk landscape (speed, accuracy, efficiency, cost effectiveness, expertise required)?
        \item Are there risks related to: identification and exploitation of vulnerabilities, spear phishing, or bug finding?
    \end{itemize} \\ \hline

\textbf{Privacy} & 
    \begin{itemize}
        \item Can the models produce or be used to surface private information?
    \end{itemize} \\ \hline

\textbf{Medicine / Healthcare} & 
    \begin{itemize}
        \item What are the capabilities and limitations of the model used in healthcare applications and contexts? Especially with regards to limitations, how do those limitations meaningfully alter the risk landscape?
        \item Do we see changes in over-confidence in answers?
        \item Do we see increased hallucination, particularly of citations?
    \end{itemize} \\ \hline

\textbf{Law} & 
    \begin{itemize}
        \item What are the capabilities and limitations of the model used in legal applications and contexts? Especially with regards to limitations, how do those limitations meaningfully alter the risk landscape?
        \item Do we see changes in over-confidence in answers?
        \item Do we see increased hallucination, particularly of citations?
    \end{itemize} \\ \hline

\textbf{Tool Use} & 
    \begin{itemize}
        \item Where does tool use (browsing, chaining with other systems, using code interpreter, etc.) meaningfully augment the model’s base capabilities in ways that might pose risks? Example areas include (carrying out purchases or financial transactions, etc.)
    \end{itemize} \\ \hline

\textbf{Dangerous Planning} & 
    \begin{itemize}
        \item How might the model aid in planning for dangerous or high-stakes activities (attack planning, obtaining or distributing illegal or dangerous materials, circumvention of safeguards or protections in various contexts)?
    \end{itemize} \\ \hline

\textbf{Politics and Election Considerations} & 
    \begin{itemize}
        \item How might the model aid in various ways to assist in political strategy, including generation of campaign materials, coming up with voter targeting approaches, analyzing and forecasting public opinion?
    \end{itemize} \\ \hline

\textbf{Bias and Fairness} & 
    \begin{itemize}
        \item Where might the model exhibit bias? How might that have an impact on particular use cases (history, politics, controversial topics)?
        \item Does the model exhibit bias based on race, ethnicity, religion, political affiliation, etc., particularly if used to make decisions in hiring, educational access, and extending credit?
    \end{itemize} \\ \hline

\textbf{CBRN Risks} & 
    \begin{itemize}
        \item How might the model impact CBRN (chemical, biological, radiological, and nuclear) threat creation? Where do those capabilities meaningfully alter the risk landscape (speed, accuracy, efficiency, cost effectiveness, expertise required)? Particular sub questions of interest:
    \end{itemize} \\ \hline

\textbf{AI Research \& Development} & 
    \begin{itemize}
        \item Can the models be used to advance methods in AI research and training? Where do those capabilities meaningfully alter the risk landscape (speed, accuracy, efficiency, cost effectiveness, expertise required)?
    \end{itemize} \\ \hline

\textbf{Situational Awareness / Autonomous Replication} & 
    \begin{itemize}
        \item Are there properties of the model that might demonstrate situational awareness or the ability to autonomously replicate?
    \end{itemize} \\ \hline

\textbf{Violence and self harm} & 
    \begin{itemize}
        \item Does the model refuse to give answers that support violence, enable self-harm, etc.?
    \end{itemize} \\ \hline

\textbf{Controversial Questions} & 
    \begin{itemize}
        \item Does the model give responses to controversial questions (social, political, ethical) that are not inclusive of different valid perspectives?
    \end{itemize} \\ \hline

\textbf{Persuasiveness} & 
    \begin{itemize}
        \item Is the model more capable in making persuasive arguments about sensitive topics?
        \item Is the model more capable or willing to form emotional attachments or give sensitive emotional/personal advice?
    \end{itemize} \\ \hline

\end{longtable}
\end{scriptsize}
\end{document}